\documentclass[aps,prl,twocolumn,amsmath,superscriptaddress,floatfix]{revtex4}
\usepackage[dvips]{graphicx}
\usepackage{amssymb}

\newcommand{\beq}{\begin{equation}}
\newcommand{\eeq}{\end{equation}}

\newcommand{\tonset}{t_\mathrm{onset}}
\newcommand{\bfb}[1]{\mbox{\boldmath $ #1 $}}
\begin{document}
\title{Impact of a viscous liquid drop}
\author{Robert D. Schroll}
\affiliation{Physics Department and the James Franck Institute, The
University of Chicago, 929 E. 57th St., Chicago, Illinois 60637}
\author{Christophe Josserand}
\author{St\'ephane Zaleski}
\affiliation{UPMC Univ Paris 06, UMR 7190,  Institut Jean Le Rond d'Alembert, F-75005 Paris, France}
\author{Wendy W. Zhang}
\affiliation{Physics Department and the James Franck Institute, The
University of Chicago, 929 E. 57th St., Chicago, Illinois 60637}
\date{\today}
\begin{abstract} 
We simulate the impact of a viscous liquid drop onto a smooth dry solid surface.  As in experiments, when ambient air effects are negligible, impact flattens the falling drop without producing a splash.  The no-slip boundary condition at the wall produces a boundary layer inside the liquid. Later, the flattening surface of the drop traces out the boundary layer.  As a result, the eventual shape of the drop is a ``pancake'' of uniform thickness except at the rim, where surface tension effects are significant.  The thickness of the pancake is simply the height where the drop surface first collides with the boundary layer. 
\end{abstract}
\maketitle

The impact of a liquid drop onto a dry solid surface lies at the heart of many important technological processes~\cite{rein93,yarin06}, from the application of a thermal spray~\cite{herman00,jiang01,fauchais04,li04,mcdonald07,molta07,shinoda07} to atomization of fuel in a combustion chamber~\cite{williams58,chigier76,lefebvre89,molta07}.  Recent experiments revealed the splash formed when a low-viscosity liquid, such as water or ethanol, first collides with a dry smooth wall at several m/s owes its existence entirely to the presence of air~\cite{xu05,quere05,xu07a, xu07}.  These results are motivating new studies on the large-scale deformations created by impact when air effects are absent as well as how a splash forms~\cite{jepsen06,duez07,mukherjee07,subramani07,deegan08,kannan08,pepper08}.  

Here we focus on the impact of a viscous liquid drop when air effects are absent.    
Recent experiments show that reducing the ambient gas pressure also suppresses the splash of a silicone oil drop. However, the form of the splash is very different. While the splash from a low-viscosity liquid develops within a few 10 $\mu$s of impact, the splash from a silicone oil develops slowly, becoming evident only after most of the liquid drop has fallen and flattened into a thin pancake~\cite{stevens07,driscoll08}.  We use an axi\-sym\-met\-ric Volume-of-Fluid (VOF) code to simulate the impact at reduced ambient pressure~\cite{lafaurie94,josserand03,josserand05}.  Our results show that a boundary layer, corresponding to a thin region where the radial flow created by impact adjusts to the no-slip condition at the wall, is created by the impact.  The boundary layer has uniform thickness.  As impact nears its end, the drop surface flattens onto the boundary layer, evolving into a pancake of uniform thickness.  

The Volume-of-Fluid simulation solves the Navier-Stokes equations, together with constraint of incompressibility, for both the liquid interior and the gas exterior at reduced pressure.  Physically appropriate boundary conditions, in particular the Laplace pressure jump across the surface due to surface tension, are enforced~\cite{lafaurie94}. In a typical run, an initially spherical liquid drop with radius $a$ collides with a dry, smooth solid surface at an impact speed $U_0$ of several m/s.  The ambient air density $\rho_g$ is kept so small that $2$-fold changes in in the value of $\rho_g$ have little effect on the liquid dynamics.  The bottom surface of the liquid drop is not broken upon impact, which corresponds to maintaining an apparent contact angle of $180^\circ$ where the liquid rim meets the wall (Fig.~\ref{fig:times}a)~\cite{wallgap}.  We require that the velocity field inside the drop satisfies both the no-flux and no-slip boundary conditions at the solid wall.

Figure~1 plots successive drop surface shapes (outlined in white) against snapshots from an experiment by Driscoll \& Nagel under the same impact conditions~\cite{driscoll08}.  We have rescaled time by the impact time-scale $\tau \equiv a/U_0$. 
The simulation agrees very well with the experiment.  After the drop hits the wall, the liquid liquid inside the drop is diverted outwards, forming a thin sheet which expands radially along the wall (Fig.~1b).  Later, the falling drop flattens, evolving towards a shape resembling a pancake with a thickened outer rim (Fig.~1c).  By $t =7.4 \tau$, a considerable time after the drop has stopped falling, the expanding drop attains its maximum extent (Fig.~1d).  After that point in time, surface tension causes the drop in the simulation to retract inwards and reform into a spherical shape, a dynamics we do not analyze. 
\begin{figure*}
\includegraphics{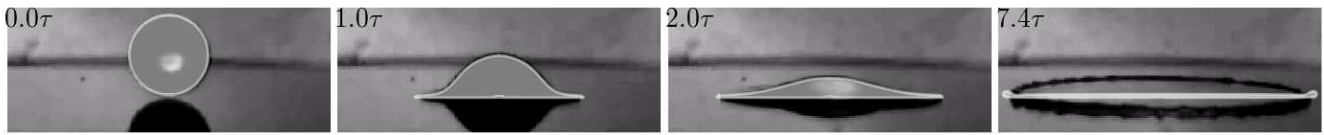} 
\caption{\label{fig:profiles} Impact of a viscous silicone oil drop at $4$ m/s onto a smooth, dry substrate at reduced ambient pressure ($34$ kPa).  Surface profiles (white) from simulated impact are overlaid against snap shots from an experiment.  From left to right, the successive times are $t=0$, $\tau$, $2\tau$ and $7.4\tau$ where $\tau \equiv a/U_0$ is the impact time-scale.  Here $Re \approx 1280$ and $We \approx 2280$. Photos courtesy of Driscoll \& Nagel.
} 
\end{figure*}

For this impact, the liquid drop has radius $a = 0.16$ cm and the impact speed $U_0$ is $4$ m/s.  The liquid is a low-molecular-weight silicone oil with dynamic viscosity $\mu_L = 9.4$ cP, density $\rho_L = 0.94$ g/cm$^3$ and surface tension $\sigma = 21$ dynes/cm.  The exterior fluid corresponds to air at $34$ kPa, with density $\rho_g =  4.4\times 10^{-4}$ g/cm$^3$ and dynamic viscosity $\mu_g = 1.8 \times 10^{-2}$ cP. The simulated impact occurs in a cylindrical tube with radius $R = 6a$ and height $H=6a$. The experiment uses a larger tube.   Changing $R$ does not change the impact results. In describing the results, we use a cylindrical coordinate system, where the tube is centered at $r=0$ and the wall lies at $z=0$. 

To correlate different impacts, we non-dimensionalize all the length-scales by $a$, the velocities by $U_0$ and the time-scales by $\tau$.   Since neither the tube dimensions nor the air properties affect the liquid dynamics reported here, the outcomes depend on only two dimensionless parameters:  the Reynolds number $Re \equiv {2 \rho_L U_0 a}/{\mu_L}$ and the Weber number $We \equiv {2 \rho_L U_0^2 a}/{\sigma}$.  The impact in Fig.~\ref{fig:profiles} correspond to $Re \approx 1280$ and $We \approx 2280$.  


How the thin liquid sheet ejected by impact evolves over time is more difficult to characterize in the experiments. Previous studies~\cite{chandra91,clanet04} proposed a simple estimate for the eventual thickness $h$ of the pancake.  The idea is that if the only mechanism that can arrest the outward radial expansion is viscous dissipation.  We then assume that the impact energy is dissipated by a radial flow of strength $U_0$ in a liquid pancake of thickness $h$ and maximal extent $R_{\rm max}$.  Balancing the dissipation against the initial kinetic energy yields
\begin{equation}
\frac{\mu_L U_0}{h^2} (\pi R^2_{\rm max}) R_{\rm max} \approx \frac{\rho_L U^2_0}{2} \left( \frac{4 \pi a^3}{3} \right) \ .
\label{eqn:Re25}
\end{equation} 
This energy balance, together with volume conservation $4 \pi a^3/3 \approx \pi R^2_{\rm max} h$, predicts that the dimensionless pancake thickness $h/a$ should scale as $Re^{-2/5}$. This feature has been confirmed by previous experiments~\cite{clanet04}.  Here we find that the eventual thickness of the pancake formed by impact from our simulation for different values of $U_0$, $\sigma$ and $\mu_L$ are entirely consistent with this estimate (see supplementary materials).

While successful, the scaling estimate for $h$ does not tell us when the characteristic pancake thickness $h$ first emerges.  Nor does it explain why the top surface of the drop evolves into a pancake of uniform thickness.  To address these questions, we examine results from the simulation. This time we switch to a reference frame $x=r-R_h(t)$.  The radial location $R_h$ corresponds to a point where the rounded rim joins onto the rest of the drop.  Within the $O(\tau)$ time window, the simulation shows that $R_h(t) \approx \sqrt{4 a U_0 t}$, so that the reference frame $x$ decelerates over time. Figure~\ref{fig:times}a displays the surface profiles at the leading edge at different moments.  Initially ($t=0.3\tau$) a thin collar is ejected from a nearly spherical drop.  As time goes on, a local minimum develops in the interface profile, separating the drop profile into a rounded rim region where surface tension effects are important and a downward sloping profile which becomes more and more gently sloped over time.  As time goes on the profile flattens to the left of the minimum.  By $t = 4.9\tau$, a pancake of uniform thickness is apparent.  At the rim, 
surface tension acts to slow the expanding liquid sheet, causing liquid to accumulate, consistent with results from previous studies~\cite{keller95,deGennes,fedorchenko05}.  

The flattening dynamics, however, has not been examined previously. To quantify its progress, we define an onset time $t_{\rm onset}$.  This is the time when the height of the interface at the local minimum first equals the eventual pancake thickness $h$. In Fig.~\ref{fig:times}b we plot $\tonset$ as a function of the impact parameters.  Non-dimensionalizing $\tonset$ by the impact time-scale $\tau = a/U_0$ produces essentially flat curves with respect to both $Re$ and $We$.  In other words, $\tonset$ is simply controlled by the kinematics of impact, with no apparent dependence on either the liquid viscosity or surface tension.  As a comparison we plot $t_{\rm max}$, the time when the drop attains its maximum extent and then retracts due to surface tension.  This quantity has a strong dependence on surface tension, or $We$.  Since $\tonset$ is considerably shorter than $t_{\rm max}$, there is still appreciable liquid motion inside the drop when the first flattening begins.  This suggests that the emergence of $h$ is not related to whether the kinetic energy has been sufficiently dissipated, but instead depends on the kinematics. 
\begin{figure}[b!]
\includegraphics{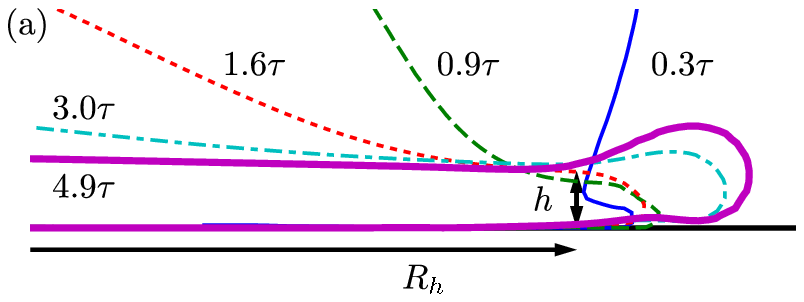}
\includegraphics{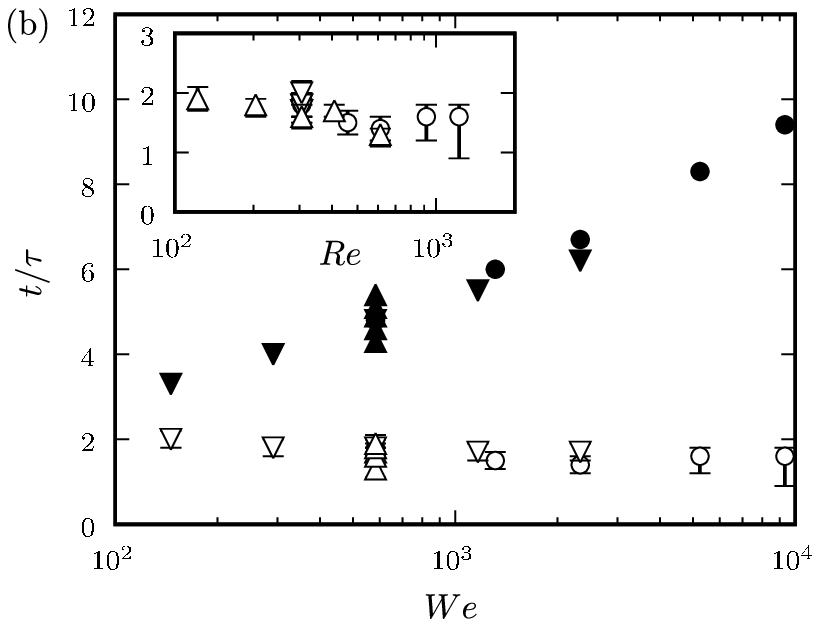}
\caption{\label{fig:times}
Time evolution of ejected liquid sheet.
(a) Leading-edge evolution in the co-moving frame $x=r-R_h(t)$ .  Here we reduced the impact speed to $U_0 = 2$ m/s and increased the liquid viscosity $\mu_L = 0.2$ poise to generate a thicker pancake.  Other parameters unchanged.
(b) Onset time $t_{\rm onset}$ (open symbols) and $t_{\rm max}$, the time of maximum extent (closed symbols) as a function of the Weber number ($We \equiv \rho_L U^2_0 a^2/\sigma$). The different symbols correspond to $U_0=2$--8 m/s ($\circ$), $\sigma=5.25$--84 dynes/cm ($\triangledown$), and $\mu=10$--50 cP ($\triangle$).  Inset plots onset time $t_{\rm onset}$ vs. $Re \equiv 2 a U_0 / \nu_L$. 
}
\end{figure}

\begin{figure}[tbp!]
\includegraphics{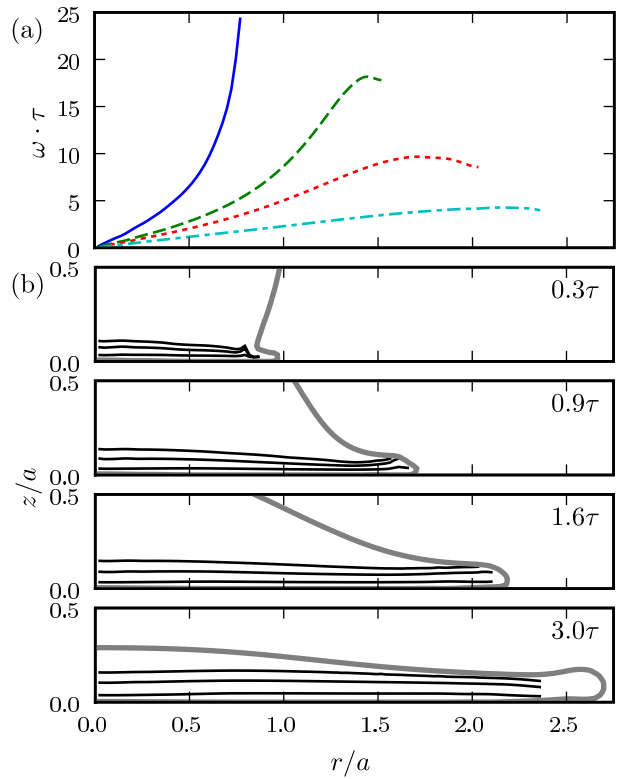}
\caption{\label{fig:vort} Vorticity evolution after impact. 
(a) Value of the vorticity at the wall as a function of radial distance $r$~\cite{vortnote}.  From top to bottom, the profiles are taken at $t=0.3 \tau$, $0.9\tau$, $1.6\tau$ and $3.0\tau$. 
(b) The 20\%, 50\%, and 90\% contours of the vorticity distribution.  }
\end{figure}

To get some insight into the kinematic origin of $h$, we examine the vorticity field.  Prior to impact, the drop is falling with a spatially-uniform downward velocity, so the vorticity is $0$ everywhere inside the drop.  After impact, the no-flux condition at the wall causes the liquid previously falling downwards to be diverted into a radially expanding flow.  This expansion flow speeds up as it moves away from the centerline, reaches a peak at $R_h$, and then slows down as it enters the rim. This radial expansion also adjusts, via viscous effects, to the no-slip boundary condition at the solid wall.  As a result of this adjustment, vorticity is generated in the liquid layer nearest to the solid wall.  At any moment, the amount of vorticity generated to ensure zero slip at the wall is dictated by the strength of the radial expansion flow. 
Because the simulated impact is axisymmetric, only the azimuthal component of the vorticity is nonzero, i.e. $\bfb\omega = \omega(r, z, t) {\bf e}_{\theta}$ where $\omega \equiv \partial u_z/\partial r - \partial u_r/\partial z$. 
For high Reynolds number flows this adjustment takes place inside a narrow boundary layer.  In Fig.~\ref{fig:vort}a we plot the wall value of the vorticity $\omega_0(r,t)$~\cite{vortnote}. Since the radial outflow is largest near the outer edge, the vorticity also has a maximum near $R_h$.  As the impact proceeds, the downward fall of the liquid drop slows, slowing the expansion and thus reducing the magnitude of $\omega_0$.

We next outline the spatial extent of the boundary layer.  This task is complicated by the fact that the absolute size of the vorticity is strongly correlated with the strength of the radial expansion flow.  Since the radial expansion flow, generated due to the no-flux condition, varies with $r$, vorticity is generated at different rates at different spatial locations. Moreover, at a given location, the vorticity production rate slows over time because the radial expansion slows.  Thus contours of absolute vorticity do not provide clear indications for the spatial extent of the boundary layer.  We side-step this complication by normalizing $\omega(r, z, t)$, the vorticity distribution in the bulk of the liquid, by the ``wall'' value $\omega_0(r, t)$. This essentially strips away variations in the vorticity distribution due to the varying speed of the radial expansion flow. For simple high Reynolds number flows, such as a uniform flow past a solid wall, as well as the boundary layer created by a straining flow towards a solid wall~\cite{acheson},
this procedure correctly reproduces the boundary layer structure that emerges from asymptotic analysis.  In Fig.~4b we plot contours of the normalized vorticity distribution.  In each snapshot, the solid lines within the liquid drop correspond to contours where $\omega(r, z, t)/\omega_0(r,t) =90 \%$ (lowest curve), $50\%$ and $20 \%$ (highest curve).  At early times ($t=0.3\tau$), the boundary layer delineated by the contours is a pancake-shaped region inside the liquid drop.  Except at the outermost edge, the top surface of the liquid drop is widely separated from the boundary layer. As impact proceeds, the boundary layer extends radially and thickens slightly, but retains its pancake shape.  At $t=\tonset$, the top surface of the drop collides with the boundary layer.  After the collision, the surface at the collision location ceases to decrease in height.  At the same time, the rest of the drop surface continues to fall downwards, bringing more and more portions of the surface into collision with the boundary layer.  The result is a ``front'' that flattens inwards radially, tracing out the pancake shaped boundary layer.  

The idea that the eventual pancake thickness $h$ first emerges when the top surface of the drop collides with the boundary layer suggests a slightly different scaling relation for the eventual thickness $h$. 
If  the boundary layer simply thickens diffusively prior to its collision with the top surface, then $h$ should scale as $\sqrt{\nu_L \tonset}$,  where $\nu_L$ is the kinematic viscosity of the liquid.  Since $\tonset$ is simply $a/U_0$ (see Fig.~2), we see that $h/a$ should obey the Blasius scaling $Re^{-1/2}$.  This is also consistent with our simulation results, basically because our range of $Re$ is too limited to resolve a scaling exponent of $1/2$ from the previously proposed $2/5$ scaling (See supplementary materials). 

Before concluding, we comment on how these results on impact without air effects may relate to splash formation when air is present. In the simulated impacts, air effects are negligible and the boundary layer always remains attached to the wall.   As $t$ approaches $2 \tau$, the radial expansion slows and the pressure gradient within the liquid is essentially zero.  In this time window, the boundary layer is not securely attached to the wall.  Any external perturbation that adds an adverse pressure gradient, \textit{e.g.}\ increased resistance from the air at larger ambient pressures, may be enough to cause the outer edge of the boundary layer to separate from the wall.  Since the surface profile is coupled to the boundary layer, the separating boundary layer may peel the thin liquid layer away from the wall, forming the beginning of a corona.  
Simulations to check this idea are underway~\cite{rdslongpaper}. 

In conclusion, we have simulated the impact of a viscous oil drop when the ambient air pressure is reduced to a very low value, so that impact at several m/s does not produce a splash.  Results on the large-scale shape deformation agree quantitatively with measurements from available experiments.  The simulation reveals that the thin, spatially uniform pancake shape that a falling drop gets flattened into owes its existence to the boundary layer in the liquid drop created by impact.

The authors are grateful to Cacey Stevens, Michelle Driscoll, and Sidney Nagel for helpful discussions and for sharing unpublished data from their experiments.  We thank Leo Kadanoff, Margo Levine, Alex Obakov, David Qu\'er\'e, and Tom Witten for encouragement and helpful comments.  This work was supported by the Keck initiative for ultrafast imaging  and a Sloan foundation fellowship (W.W.Z.). 

\bibliography{dropimpact}

\begin{thebibliography}{37}
\bibitem[{wal()}]{wallgap}
\bibinfo{note}{In the simulation, a small amount of air is trapped between the
  liquid drop and the solid wall. The amount of air entrapped is too small to
  be adequately resolved by our simulation, which uses a uniform spatial grid.
  We observe that, over the duration of the impact, most of the entrapped air
  forms into an air pocket beneath the center of the drop, leaving a very thin
  film of air separating the liquid and the wall. In a typical simulation the
  air pocket is at most tens of $\mu$m near the center. We have checked that
  changing the details of the air entrapment dynamics does not change the
  results on the liquid dynamics reported here~\cite{rdslongpaper}. Since we
  are only interested in the flow dynamics when air flow effects are
  negligible, we require that the entrapped air does not lubricate the liquid
  flow and impose the no-slip boundary condition on the liquid flow.}
\bibitem[{vor()}]{vortnote}
\bibinfo{note}{Since the bottom surface of the drop is not broken by impact, no
  liquid is ever in direct contact with the wall. We therefore define the wall
  value by taking the derivatives near the bottom surface at a fixed vertical
  height. Here we chose $z/a = 1/60$. Choosing different values makes no
  difference provided the vertical height is suitably small.}
\end{thebibliography}
\end{document}